\documentstyle[twoside,fleqn,espcrc2,epsfig]{article}

\newcommand{\lbl}[1]{\label{eq:#1}}
\newcommand{ \rf}[1]{(\ref{eq:#1})}

\newcommand{\be}{\begin{equation}}
\newcommand{\ee}{\end{equation}}
\newcommand{\bea}{\begin{eqnarray}}
\newcommand{\eea}{\end{eqnarray}}
\newcommand{\setl}{\setlength\arraycolsep{2pt}}

\newcommand{\noi}{\noindent}
\newcommand{\nn}{\nonumber}
\newcommand{\ra}{\rightarrow}

\newcommand{\lesssim}{ {\
\lower-1.2pt\vbox{\hbox{\rlap{$<$}\lower5pt\vbox{\hbox{$\sim$}}}}\ } 
}
\newcommand{\gtrsim}{ {\
\lower-1.2pt\vbox{\hbox{\rlap{$>$}\lower5pt\vbox{\hbox{$\sim$}}}}\ } 
}

\newcommand{\cO}{{\cal O}}

\newcommand{\Imm}{\mbox{\rm Im}}

\newcommand{\MeV}{\mbox{\rm MeV}}
\newcommand{\GeV}{\mbox{\rm GeV}}

\newcommand{\annd}{\mbox{\rm and}}

\newcommand{\als}{\alpha_{\mbox{\rm {\scriptsize s}}}}

\newcommand{\exxp}{\mbox{\rm \tiny exp.}}

\newcommand{\NQCD}{QCD$_{\infty}~$}
\newcommand{\thh}{\mbox{\rm {\tiny th}}}

\input epsf

\newcommand{\AmS}{{\protect\the\textfont2
  A\kern-.1667em\lower.5ex\hbox{M}\kern-.125emS}}

\title{Minimal Hadronic Ansatz to Large $N_c$ QCD and Hadronic
$\tau$--Decay\thanks{
    Work supported in part by TMR, EC-Contract
    No. ERBFMRX-CT980169(EURODA$\phi$NE).}}

\author{Eduardo de Rafael\address{Centre de Physique Th{\'e}orique,
        CNRS-Luminy, Case 907, \\
        F-13288 Marseille Cedex 9, France}}

\begin{document}

\begin{abstract}

I report on some recent work done in collaboration with Santi
Peris and Boris Phily~\cite{PPdeR00} where, using the Aleph data on vector and
axial-vector spectral functions, we test simple duality properties of QCD in
the large $N_c$ limit which emerge in the approximation of a {\it minimal
hadronic ansatz} of a spectrum of narrow states. These duality properties
relate the short--  and long--distance behaviours of specific correlation
functions, which are order parameters of spontaneous chiral symmetry
breaking, in a  way that we find well supported by the data..

\end{abstract}

% typeset front matter (including abstract)
\maketitle

\section{INTRODUCTION}

At first sight, the {\it hadronic world} predicted by QCD in the
limit of a large number of colours $N_c$~\cite{tH74} may seem rather
different from the real world. The hadronic spectrum of vector and
axial--vector states, observed e.g. in $e^{+} e^{-}$ annihilations and in
$\tau$ decays, has certainly much more structure than the infinite set of
narrow states predicted by large $N_c$ QCD~\cite{Wi79} (\NQCD). There are,
however, many instances in Particle Physics where one is only
interested in certain weighted integrals of hadronic spectral
functions. In these cases, it may be enough to know a few {\it global}
properties of the hadronic spectrum; one does not expect the integrals
to depend crucially on the details of
the spectrum at all energies. Typical examples of that are the coupling
constants of the effective chiral Lagrangian of
QCD at low energies, as well as the coupling constants of
the effective chiral Lagrangian of the electroweak interactions of
pseudoscalar particles in the Standard Model, which are needed to
understand
$K$--Physics in particular, (see e.g. the review article in
ref.~\cite{Pi99} and references therein.) It is in these examples that the
{\it hadronic world} predicted by \NQCD may provide a good approximation to
the real hadronic spectrum. If so,
\NQCD could then become a useful phenomenological approach for understanding
non--perturbative QCD physics at low energies.

There are indeed a number of successful
calculations which have already been done
within the framework of \NQCD, (see
ref.~\cite{KPdeR98,KPPdeR99,KPdeR99,KPdeR00,PdeR00} and references therein.)
The picture which emerges from these applications is one of a remarkable
simplicity. It is found that, when dealing with Green's functions that are
{\it order parameters} of spontaneous chiral symmetry breaking, the
restriction of the infinite set of large $N_c$ narrow states to a {\it
minimal hadronic ansatz} which is needed to satisfy the leading short-- and
long--distance behaviours of the relevant Green's functions, provides already
a very good approximation to the observables one computes. The purpose of the
work in ref.~\cite{PPdeR00}, which I am reporting here, is to investigate
this {\it minimal hadronic ansatz} approximation in a case where one can
compare, in detail, the theoretical predictions to the phenomenological
results evaluated with experimental data.
%%%%%%%%%%%%%%%%%%%%%%%%%%%%%%%%%%%%%%%%%%%%%%

\section{THE LEFT--RIGHT CORRELATION FUNCTION}

Of particular interest for our purposes is the correlation 
function ($Q^2\equiv -q^2\ge 0$ for $q^2$ space--like)
\be\lbl{lrtpf}
\Pi_{LR}^{\mu\nu}(q)= 
 2i\int d^4 x\,e^{iq\cdot x}\langle 0\vert
\mbox{\rm T}\left(L^{\mu}(x)R^{\nu}(0)^{\dagger}
\right)\vert 0\rangle\,,
\ee 
with colour singlet currents
\be
R^{\mu}\left(L^{\mu}\right)=
\bar{d}(x)\gamma^{\mu}\frac{1}{2}(1\pm\gamma_{5})u(x)\,.
\ee 
In the chiral limit, ($m_{u,d,s}\ra 0$\,,) this correlation function has
only a transverse component
\be\lbl{lritpf}
\Pi_{LR}^{\mu\nu}(Q^2)=(q^{\mu}q^{\nu}-g^{\mu\nu}q^2)\Pi_{LR}(Q^2)\,.
\ee
The self--energy like function $\Pi_{LR}(Q^2)$  vanishes order by order
in  perturbative QCD (pQCD) and
is an order parameter of S$\chi$SB for all values of $Q^2$;
therefore it obeys an unsubtracted dispersion relation
\be\lbl{disprel}
\Pi_{LR}(Q^2)=\int_{0}^{\infty}dt\frac{1}{t+Q^2}
\frac{1}{\pi}\Imm\Pi_{LR}(t)\,.
\ee

In \NQCD the spectral function $\frac{1}{\pi}\Imm\Pi_{LR}(t)$ consists 
of
the difference of an infinite number of narrow vector and axial--vector
states, together with the Goldstone pole of the pion:
\bea
\lefteqn{\frac{1}{\pi}\Imm\Pi_{LR}(t) =   
\sum_{V}f_{V}^2 M_{V}^2\delta(t-M_{V}^2)} \nn \\
 & & -F_{0}^2\delta(t) -\sum_{A}f_{A}^2
M_{A}^2\delta(t-M_{A}^2)\,.
\eea 
The low $Q^2$ behaviour of $\Pi_{LR}(Q^2)$, i.e.
the long--distance behaviour of the correlation function in
Eq.~\rf{lrtpf}, is governed by chiral perturbation theory:
\be 
-Q^2\Pi_{LR}(Q^2)\vert_{Q^2\ra 0}=F_{0}^2\!+\!4L_{10}Q^2\!+\!\cO(Q^4)\,,
\ee 
where $F_{0}$ is the pion coupling constant in the chiral limit, and
$L_{10}$ is one of the coupling constants
of the $\cO(p^4)$ effective chiral Lagrangian. 
The high $Q^2$
behaviour of $\Pi_{LR}(Q^2)$, i.e.
the short--distance behaviour of the correlation function in
Eq.~\rf{lrtpf}, is governed by the operator product expansion (OPE) 
of the
two local currents in Eq.~\rf{lrtpf}~\cite{SVZ79},
\be\lbl{OPE}
\lim_{Q^2\ra\infty}\!\!Q^6\Pi_{LR}(Q^2)=\!\! 
 \left[-4\pi^2\frac{\alpha_s}{\pi}\!+\!\cO(\alpha_s^2)\right]
\!\!\langle\bar{\psi}\psi\rangle^2\,,
\ee
which implies the two Weinberg sum rules:
\be\lbl{weinbergsr1}
\int_{0}^{\infty}\!\!\!\!dt\Imm\Pi_{LR}(t)\!\!=\!\!\sum_{V}\!f_{V}^2
\!M_{V}^2-\!\!\sum_{A}\!f_{A}^2 \!M_{A}^2\!\!-\!\!F_{0}^2\!\!=\!\!0\,,
\ee
and
\be\lbl{weinbergsr2}
\int_{0}^{\infty}\!\!\!\!dt t\Imm\Pi_{LR}(t)\!=\!\sum_{V}f_{V}^2
M_{V}^4\!-\!\!\sum_{A}f_{A}^2 M_{A}^4\!=\!0\,.
\ee 

In fact, as pointed out in ref.~\cite{KdeR98}, in \NQCD, there exist an
infinite number of Weinberg--like sum rules. In full
generality, the moments of the
$\Pi_{LR}$ spectral function with $n=3,4,\dots$, 
\bea\lbl{posmoments}
\lefteqn{
\int_{0}^{\infty} dt\,t^{n-1}\left[\frac{1}{\pi}\Imm\Pi_{V}(t)-
\frac{1}{\pi}\Imm\Pi_{A}(t)\right]=} \nn \\
 & & \sum_{V} f_{V}^2 M_{V}^{2n} - \sum_{A} f_{A}^2 M_{A}^{2n}\,,
\eea
govern the short--distance expansion of the $\Pi_{LR}(Q^2)$
function
\bea\lbl{largeQ}
\lefteqn{
\Pi_{LR}(Q^2\!)\vert_{Q^2\ra\infty}\!\!=\! 
\left(\!\sum_{V} f_{V}^2 M_{V}^6 \!- \!\!\sum_{A} f_{A}^2
M_{A}^6\!\right)\!\!\frac{1}{Q^6}} \nn
\\ &  & + \left(\sum_{V} f_{V}^2 M_{V}^8 - \sum_{A} f_{A}^2
M_{A}^8\right)\frac{1}{Q^8} +
\cdots \,.
\eea
On the other hand, inverse moments of the
$\Pi_{LR}$ spectral function, with the pion pole removed, (which we denote
by
$\Imm\tilde{\Pi}_{A}(t)$,) determine a class of  
coupling constants
of the low--energy effective chiral Lagrangian. For example,
\bea\lbl{invmom}
\lefteqn{
\int_{0}^{\infty} dt \frac{1}{t}\left[\frac{1}{\pi}\Imm\Pi_{V}(t)-
\frac{1}{\pi}\Imm\tilde{\Pi}_{A}(t)\right]=} \nn \\
 & & \sum_{V} f_{V}^2 -\sum_{A} f_{A}^2 = -4 L_{10}\,.
\eea
Moments with higher inverse powers of $t$ are associated with 
couplings of
composite operators of higher dimension in the chiral Lagrangian. Tests of
the two Weinberg sum rules in Eqs.~\rf{weinbergsr1} and \rf{weinbergsr2}
and of the
$L_{10}$ sum rule in Eq.~\rf{invmom}, in a different context to the one we
are interested in here, have often appeared in the literature, (see e.g.
refs.~\cite{DHGS98} and \cite{DS99} for recent discussions where earlier
references can also be found.)
%%%%%%%%%%%%%%%%%%%%%%%%%%%%%%%%%%%%%%%%%%%%%%
%%%%%%%%%%%%%%%%%%%%%%%%%%%%%%%%%%%%%%%%%%%%%%

\section{THE MINIMAL ANSATZ}

We shall now consider the approximation which we call the {\it
minimal hadronic ansatz} to \NQCD. In the case of the left--right two--point
function in Eq.~\rf{lrtpf}, this is the approximation where the hadronic
spectrum consists of one vector state $V$, one axial--vector state
$A$ and the Goldstone pion, with the ordering~\cite{KdeR98} $M_{V}< M_{A}$.
This is the {\it minimal spectrum} which is required  to satisfy the two
Weinberg sum rules in Eqs.~\rf{weinbergsr1} and
\rf{weinbergsr2}.  
In this approximation,
$\Pi_{LR}(Q^2)$ has a very simple form

\bea\lefteqn{
-Q^2
 \Pi_{LR}(Q^2) =  
\frac{F_{0}^2}{\left(1+\frac{Q^2}{M_{V}^2}\right)
\left(1+\frac{Q^2}{M_{A}^2}\right)}}  \\
& & \ \ \    = \frac{M_{A}^2
M_{V}^2}{Q^4}\frac{F_{0}^2}{\left(1+\frac{M_{V}^2}{Q^2}\right)
\left(1+\frac{M_{A}^2}{Q^2}\right)}\,.
\eea

\noi 
This equation shows, explicitly, a remarkable short--distance
$\rightleftharpoons$ long--distance duality~\cite{deR99}. Indeed, 
with $g_{A}$ defined
so that $M_{V}^2\!\!=\!\!g_{A}M_{A}^2$ and
$z\!\!\equiv\!\!\frac{Q^2}{M_{V}^2}$,  the non--local order
parameters corresponding to the long--distance expansion for $z\ra 0$,
which are couplings of the effective chiral Lagrangian i.e., 
\bea\lbl{chiralex}
\lefteqn{ 
-Q^2\Pi_{LR}(Q^2)\vert_{z\ra 0}=
F_{0}^2\left\{1-(1+g_{A})z\right.} \nn \\
 & & \left. +(1+g_{A}+g_{A}^2)z^2+\cdots\right\}\,,
\eea 
are correlated to the  local order parameters of the
short--distance OPE for $z\ra\infty$ in a very simple
way:
\bea\lbl{opemha}
\lefteqn{-Q^2
\Pi_{LR}(Q^2)\vert_{z\ra\infty}\!=\!
F_{0}^2\frac{1}{g_{A}}\frac{1}{z^2}
\left\{\!1\!-\! \left(\!1\!+\!\frac{1}{g_{A}}\right)\frac{1}{z}\right.} \nn \\
& & \left. +
\left(1+\frac{1}{g_{A}}+\frac{1}{g_{A}^2}\right)\frac{1}{z^2}
+\cdots\right\}\,;
\eea
in other words, there is a one to one correspondence 
between the two expansions
by changing
\be
g_{A}\rightleftharpoons \frac{1}{g_{A}}\quad \annd \quad
z^{n}\rightleftharpoons \frac{1}{g_{A}}\frac{1}{z^{n+2}}\,.
\ee
The moments of the $\Pi_{LR}$ spectral function, when evaluated in
the  {\it minimal hadronic ansatz} approximation, can be converted
into a very simple set of finite energy sum rules (FESR's), corresponding
to the OPE in Eq.~\rf{opemha}  
{\setl
\bea\lbl{moments2}
\int_{0}^{s_0}
\!\!\!dt\,t^{2}\frac{1}{\pi}\Imm\Pi_{LR}(t) & \!=\!
&-F_{0}^{2}M_{V}^4\frac{1}{g_{A}}\,,
\\ \lbl{moments3}
\int_{0}^{s_0}
\!\!dt\,t^{3}\frac{1}{\pi}\Imm\Pi_{LR}(t) & \!=\!
&-F_{0}^{2}M_{V}^6\frac{1\!+\!\frac{1}{g_{A}}}{g_{A}}\,, \\
\lbl{moments4}
\int_{0}^{s_0}
\!\!dt\,t^{4}\frac{1}{\pi}\Imm\Pi_{LR}(t) & \!=\!
&-F_{0}^{2}M_{V}^8\frac{1\!+\!\frac{1}{g_{A}}\!+\!\frac{1}{g_{A}^2}}{g_{A}}\,,
\\
\cdots\qquad\qquad &  &\qquad\cdots\,\,. \nn
\eea}

\vspace{-0.5cm}
\noi
where the upper limit of integration $s_{0}$ denotes the onset
of the pQCD continuum which, in the chiral limit, is common to the
vector and axial--vector spectral functions. It is important to realize
that $s_{0}$ is not a free parameter. Its value is fixed by the requirement
that the OPE of the correlation function of two vector currents, (or two
axial--vector currents,) in the chiral limit, have no
$1/Q^2$ term, which results in an implicit equation for
$s_{0}$~\cite{BLdeR85,PPdeR98}. In the
{\it minimal hadronic ansatz} approximation the onset of the pQCD continuum,
which we shall call
$s_{0}^{*}$, is then fixed by the equation
\be\lbl{duality}
\frac{N_c}{16\pi^2}\frac{2}{3}s_{0}^{*}\left(1+\cO(\als)\right)=F_{0}^2
\frac{1}{1-g_{A}}\,.
\ee
Also, the moments which
correspond to the chiral expansion in Eq.~\rf{chiralex} are given by
another simple set of FESR's: {\setl
\bea\lbl{inversemoments}
\int_{0}^{s_0}
dt\,\frac{1}{\pi}\Imm\tilde{\Pi}_{LR}(t) & =
& F_{0}^2\,,
\\
\lbl{inversemoments2}
\int_{0}^{s_0}
\frac{dt}{t}\,\frac{1}{\pi}\Imm\tilde{\Pi}_{LR}(t) & =
& \frac{F_{0}^2}{M_{V}^2}(1\!+\!g_{A})\,, \\
\lbl{inversemoments3}
\int_{0}^{s_0}
\frac{dt}{t^2}\,\frac{1}{\pi}\Imm\tilde{\Pi}_{LR}(t) & =
& \frac{F_{0}^2}{M_{V}^4}(1\!+\!g_{A}\!+\!g_{A}^2)\,,
\\
\cdots\qquad\qquad &  &\qquad\cdots\,. \nn
\eea}

\vspace{-0.5cm}
\noi  
We propose to test these duality relations by comparing  moments of
the \underline{physical} spectral function
$\frac{1}{\pi}\Imm\Pi_{LR}^{\exp}(t)$ determined from experiment to the
predictions of the {\it minimal hadronic ansatz} as shown in the r.h.s. of
Eqs.~\rf{moments2} to
\rf{moments4} and  Eqs.~\rf{inversemoments} to \rf{inversemoments3}.
%%%%%%%%%%%%%%%%%%%%%%%%%%%%%%%%%%%%%%%%%%%%%%%%%%%%%%%%%%%%%%%%%%%%
%%%%%%%%%%%%%%%%%%%%%%%%%%%%%%%%%%%%%%%%%%%%%%%%%%%%%%%%%%%%%%%%%%%%
%%%%%%%%%%%%%%%%%%%%%%%%%%%%%%%%%%%%%%%%%%%%%%%%%%%%%%%%%%%%%%%%%%%%

\section{EXPERIMENTAL MOMENTS VERSUS THE PREDICTIONS OF THE MINIMAL
HADRONIC ANSATZ APPROXIMATION TO \NQCD  }

The ALEPH collaboration at LEP has measured the
inclusive invariant mass--squared distribution of hadronic $\tau$
decays~\cite{ALEPH} into non--strange particles. They have been able to
extract from their data, both, the vector current spectral function
$\frac{1}{\pi}\Imm\Pi_{V}^{\exxp}(t)$ and the axial--vector current
spectral function
$\frac{1}{\pi}\Imm\Pi_{A}^{\exxp}(t)$ up to
$t\simeq 3\,\GeV^2$. In fact, in the real world, the correlation
function in Eq.~\rf{lritpf} has a non--transverse term as well,
which is dominated by the pion pole contribution to the axial--vector
component. The vector contribution to the non--transverse term vanishes in
the limit of isospin invariance.

In order to compare the moments of the experimental  spectral
function
$\frac{1}{\pi}\Imm\Pi_{LR}^{\exxp}(t)$ to those in Eqs.~\rf{moments2}
to
\rf{moments4} and  Eqs.~\rf{inversemoments} to \rf{inversemoments3} we still 
have to correct for the fact that the
FESR's in these equations apply in the chiral limit where
$m_{u,d}\ra 0$. This we do by exploiting the analyticity properties of
the two--point function $\Pi_{LR}$ in the complex $q^2$--plane.
Integration over a standard contour relates weighted integrals of the
spectral function $\frac{1}{\pi}\Imm\Pi_{LR}^{\exxp}(t)$ in a finite
interval on the real axis to integrals of
$\Pi_{LR}(q^2)$ over a {\it small} circle
$\vert q^2\vert=s_{\thh}$ and a {\it large} circle $\vert q^2\vert=s_{0}$:
\bea\lbl{contour}
\lefteqn{\int_{s_{th}}^{s_0} dt f(t) \Imm \Pi_{LR} (t) 
=\!\!\!\!\!\oint\limits_{|q^2|=s_{th}}\!\!\!\!\! dq^2 \frac{1}{2i}f(q^2)
\Pi_{LR} (q^2)} \nn \\ & &  
 -
\!\!\!\oint\limits_{|q^2|=s_0}\!\!\!\!\! dq^2 \frac{1}{2i}f(q^2)
\Pi_{LR} (q^2)\,,
\eea

\noi
where the weight function $f(q^2)$ is a conveniently chosen analytic
function inside the contour; in our case simple powers and inverse powers
of $q^2$. The chiral corrections in the {\it small} circle are particularly
important in the evaluation of the inverse moments. We have evaluated them
by taking into account the one loop expression of
$\Pi_{LR}(z)$ in chiral perturbation theory~\cite{GL84}. The chiral
corrections in the {\it large} circle are rather small. They appear as
leading $1/Q^2$ and next--to--leading
$1/Q^4$ power corrections in the OPE of $\Pi_{LR}(Q^2)$ at large
$Q^2$ but their coefficients, proportional to quark masses, are
small~\cite{FNdeR79}. With these corrections incorporated, we proceed
now to the comparison we are looking for. This is shown in Figs.~1 and 2
below where we show the various moments as a function of $s_{0}$.
The three plots in Fig.~1 
show the experimental
moments on the l.h.s. of Eqs.~\rf{moments2},
\rf{moments3} and  \rf{moments4} as a function of $s_{0}$, extrapolated at the
chiral limit as discussed above, and normalized to the corresponding {\it
minimal hadronic ansatz} predictions given on the r.h.s. of these equations.

%%%%%%%%%%%%%%%%%%%%%%%%%%%%%%%%%%% 
\centerline{\epsfbox{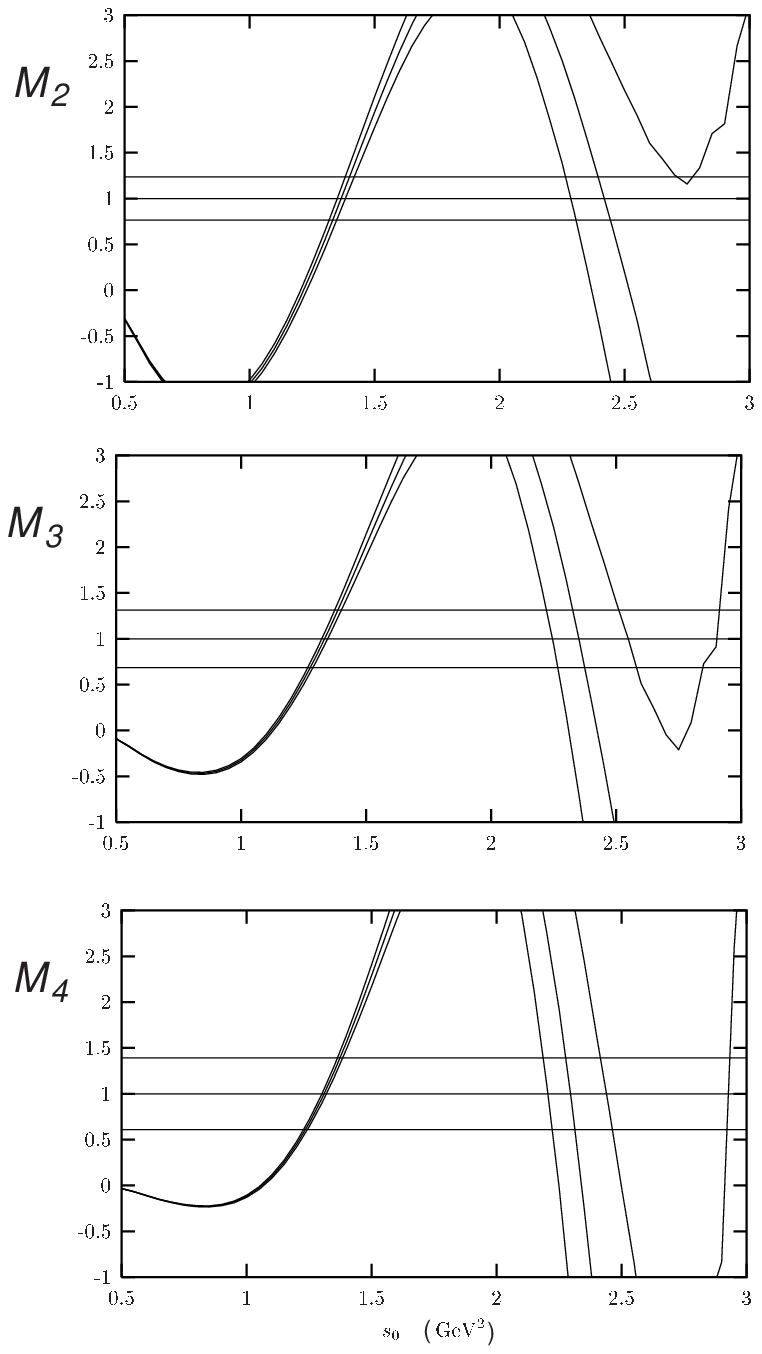}}
\vspace{0.3cm}
{\bf{Fig.~1}} {\it Plot of the experimental moments in Eqs.~\rf{M2}, 
\rf{M3} and \rf{M4} normalized to the 
{\it minimal hadronic ansatz}
predictions.}
\vspace{0.25cm}
%%%%%%%%%%%%%%%%%%%%%%%%%%%%%%%%%%

\noi
The three curves $M_{2}$, $M_{3}$ and $M_{4}$ in Fig.~1 correspond to the
quantities: {\setl
\bea\lbl{M2}
M_{2}&\!\!=&\!\!\frac{-g_{A}}{F_{0}^2M_{V}^4}\!\!\int_{0}^{s_{0}}\!\!\!\!dt
t^2\frac{1}{\pi}\Imm\Pi_{LR}^{\exp}(\!t\!)\,, \\
\lbl{M3} M_{3}&\!\!=&\!\!\frac{-g_{A}}{F_{0}^2M_{V}^6\left(1+\frac{1}{g_{A}}
\right)}\!\!\int_{0}^{s_{0}}\!\!\!\!dt
t^3\frac{1}{\pi}\Imm\Pi_{LR}^{\exp}(t)\,, \\
\lbl{M4}M_{4}&\!\!=&\!\!\frac{-g_{A}}{F_{0}^2M_{V}^8\!\left(\!1\!+\!\frac{1}{g_{A}}\!
+\!\frac{1}{g_{A}^2}
\!\right)}\!\!\int_{0}^{s_{0}}\!\!\!\!dt
t^3\frac{1}{\pi}\Imm\Pi_{LR}^{\exp}(\!t\!)\,.
\eea}

\noi
On the other hand, the three plots in Fig.~2 
show the experimental inverse
moments on the l.h.s. of Eqs.~\rf{inversemoments},
\rf{inversemoments2} and  \rf{inversemoments3} as a function of $s_{0}$, 
with the pion pole removed, extrapolated at the chiral limit as
discussed above, and normalized to the corresponding {\it minimal hadronic
ansatz} predictions given on the r.h.s. of these equations. The three curves
$M_{0}$, $M_{-1}$ and $M_{-2}$ in Fig.~2 correspond to the quantities:
{\setl
\bea
\lbl{M0}M_{0}&\!\!=&\!\!\frac{1}{F_{0}^2}\!\!\int_{0}^{s_{0}}\!dt
\frac{1}{\pi}\Imm\tilde{\Pi}_{LR}^{\exp}(\!t\!)\,, \\
\lbl{MM1}M_{-1}&\!\!=&\!\!\frac{M_{V}^2}{F_{0}^2\left(1+g_{A}
\right)}\!\!\int_{0}^{s_{0}}\!\frac{dt}{t}\frac{1}{\pi}\Imm\tilde{\Pi}_{LR}^{\exp}(t)\,,
\\
\lbl{MM2}M_{-2}&\!\!=&\!\!\frac{M_{V}^4}{F_{0}^2\!\left(\!1\!+\!g_{A}\!
+\!g_{A}^2
\!\right)}\!\!\int_{0}^{s_{0}}\!\frac{dt}{t^2}
\frac{1}{\pi}\Imm\tilde{\Pi}_{LR}^{\exp}(\!t\!)\,.
\eea}

\noi
The horizontal bands on
the plots in Figs.~1 and 2 correspond to the induced error of the {\it minimal
hadronic ansatz} predictions
from the input values:
$F_{0}\!=\!87\pm3.5\,\MeV$, $M_{V}\!=\!748\pm29\,\MeV$ and $g_{A}\!=\!0.50\pm
0.06$. These are the values favored by a global fit of the {\it minimal
hadronic ansatz} to low--energy observables~\cite{PPdeR98}. The moments
$M_n$, with the experimental error propagation included, are the curved
bands in the figures.

The remarkable feature which the curves in Figs.~1 and 2 show is that,
within errors, there is a crossing of \underline{all} the experimental
moments with the {\it minimal hadronic ansatz} band which takes place in the
\underline{same}
$s_{0}$ region, i.e., around $s_{0}\sim 1.4~\GeV^2$, rather close indeed to
the
$s_{0}^{*}$ value which follows from the duality relation in Eq.~\rf{duality}:
$s_{0}^{*}\!\!=(1.2\pm 0.2)~\GeV^2$. The same happens for the 2nd
Weinberg sum rule in Eq.~\rf{weinbergsr2}, which we show in Fig.~3, where
\be\lbl{M1}
M_{1}\!\!=\!\!\frac{1}{F_{0}^2 M_{V}^2}\int_{0}^{s_{0}}\!dt t
\frac{1}{\pi}\Imm\tilde{\Pi}_{LR}^{\exp}(\!t\!)\,.
\ee
 The
1st Weinberg sum rule is the equivalent of the moment $M_{0}$ in
Eq.~\rf{M0}.  In fact, the agreement for the inverse moments is excellent.
This is due to the fact that inverse moments put more and more weight on the
low energy tail of the spectral function, which is known to be dominated by
the
$\rho$--resonance.

%%%%%%%%%%%%%%%%%%%%%%%%%%%%%%%%%%% 
\vspace{0.25cm}
\centerline{\epsfbox{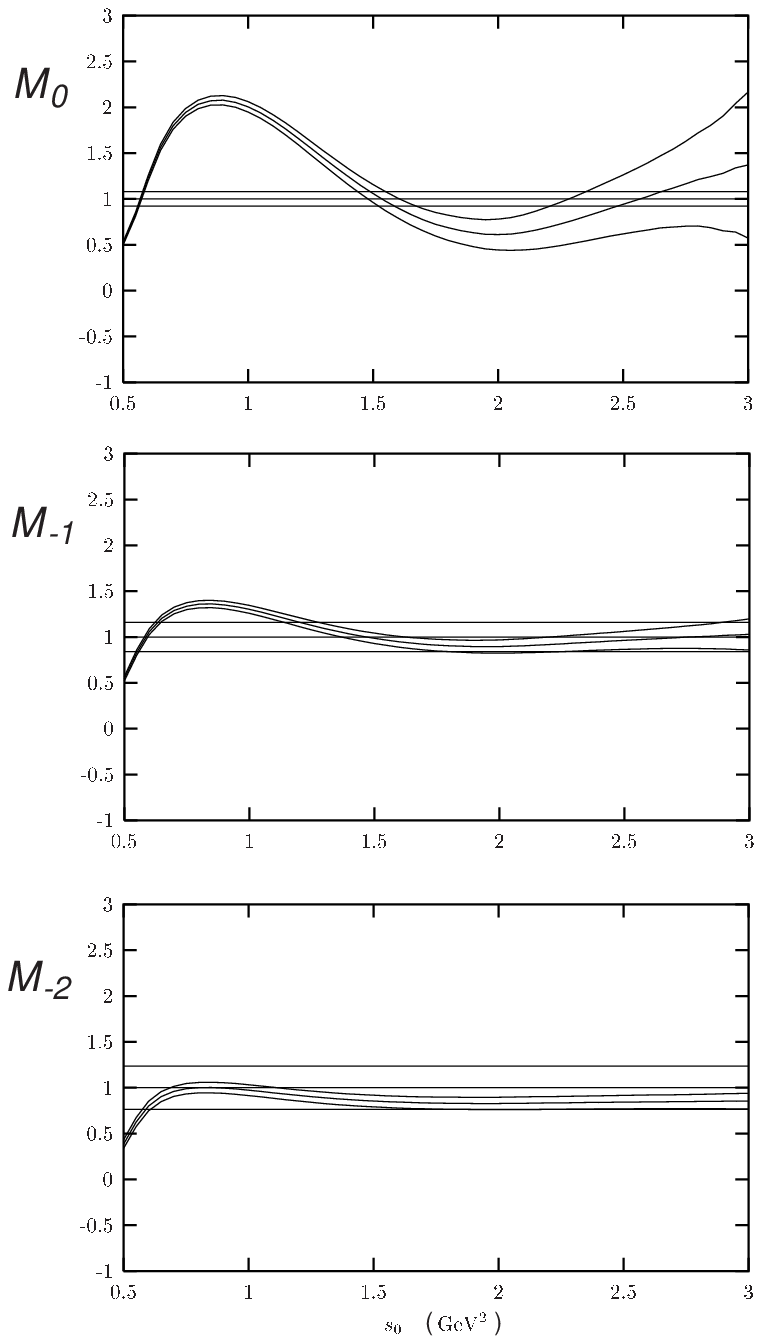}}
\vspace{0.3cm}
{\bf{Fig.~2}} {\it Plot of the experimental moments in Eqs.~\rf{M0},
\rf{MM1} and  \rf{MM2} normalized to the 
{\it minimal hadronic ansatz}
predictions.}
\vspace{0.25cm}
%%%%%%%%%%%%%%%%%%%%%%%%%%%%%%%%%%

\noi
By contrast, the
positive moments are very sensitive to the cancellations between opposite
parity hadronic states; this is why the experimental curves show larger
and larger oscillations as one increases the power of the moment.
In spite
of that, it is quite impressive that, when restricted to the $s_{0}$ region
of duality, the experimental moments agree well with the {\it minimal
hadronic ansatz} prediction, even for rather large powers which correspond
to vacuum expectation values of operators of higher and higher dimension. 

%%%%%%%%%%%%%%%%%%%%%%%%%%%%%%%%%%% 
\vspace{0.50cm}
\centerline{\epsfbox{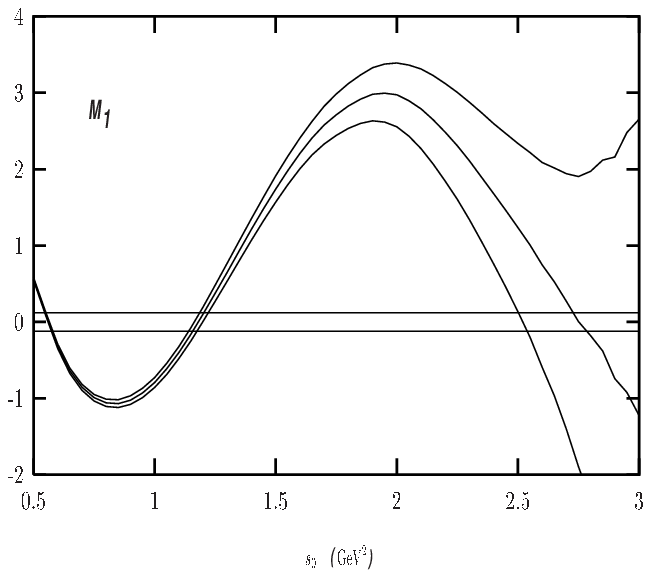}}
\vspace{0.3cm}
{\bf{Fig.~3}} {\it Plot of the 2nd Weinberg sum rule in Eq.~\rf{M1}.}
\vspace{0.25cm}
%%%%%%%%%%%%%%%%%%%%%%%%%%%%%%%%%%

\noi
We conclude that the experimental data from ALEPH is consistent with the
simple pattern of duality properties between short-- and
long--distances which follow from the  {\it minimal hadronic ansatz} of a
narrow vector and a narrow axial-vector states plus the Goldstone
pion in large--$N_c$ QCD. At the phenomenological level, it would be very
interesting to see what the impact of the choice of the upper limit $s_{0}$
is in the empirical determination of the QCD condensates, when $s_{0}$ is
restricted to the duality region.

\vskip 0.5cm
\noi
{\bf Acknowledgements}

\noi
I wish to thank Santi Peris and Boris Phily for a very
pleasant  collaboration.  I also wish to thank Stephan Narison for
organizing this interesting series of QCD conferences.

\vskip 0.3cm
\noi
{\bf QUESTIONS}
\vskip 0.25cm

\noi
{\bf S.~Narison}, Montpellier

\noi
{\it When you compare your large--$N_c$ QCD approximation with the
$\tau$--data, you find two solutions for $s_{0}$ and you choose the lowest
value. Can you explain the reason for that?}

\noi
{\bf E.~de Rafael}, CPT-Marseille

\noi
This is a good question. I did not go through that because of time
limitations. What you call the first solution, which I agree it is the one we
consider, corresponds to the {\it minimal hadronic ansatz} which I discussed.
Higher solutions correspond, very likely, to a more elaborated choice of the
spectrum: e.g., there is a second $s_{0}$ of duality if we also
include another V--sate, and the data seems to indicate that. (The detailed
analyses should appear in Boris Phily's thesis.)

\vskip 0.25cm

\noi
{\bf H.~Fritzsch}, M\"{u}nchen

\noi
{\it I am not surprised that your minimal ansatz works so well. In 1974,
Leutwyler and I wrote a paper studying such an ansatz and showing that the
coupling strengths of the lowest vector mesons fix the number of colors to be
three.}

\noi
{\bf E.~de Rafael}, CPT-Marseille

\noi
Yes, vector meson dominance is a good old idea which goes back to early
work by Sakurai. What we are doing now is to show how some of its
phenomenological good features are now naturally incorporated within the
framework of QCD at large--$N_c$. Technically, in our language, the
equation which you probably considered should be essentially the same as
our duality equation in
\rf{duality}. You can certainly use this equation to fix $N_c$, provided you
make  {\it an a priori guess} of the onset of the continuum $s_{0}$. You
probably took
$s_{0}\sim 1~\GeV$ on phenomenological grounds and got $N_c\sim 3$.   

\end{document}